\documentclass{article}
\usepackage{epsfig,amssymb,graphicx}
\newcommand{\dd}{{\rm d}}
\newcommand{\NN}{{\mathcal{N}}}
\begin{document}
\title{Predicting spike times of a detailed conductance-based neuron model driven by stochastic spike arrival}
\author{Renaud Jolivet\footnote{Address for correspondence: renaud.jolivet$@$epfl.ch}\enspace\ and Wulfram Gerstner \vspace{.5cm}\ \\
\small{Laboratory of Computational Neuroscience}\\
\small{Brain Mind Institute - I$\&$C - EPFL}}
\date{}
\maketitle
\begin{abstract}
\hspace{-\parindent}Reduced models of neuronal activity such as
Integrate-and-Fire models allow a description of neuronal
dynamics in simple, intuitive terms and are easy to simulate
numerically. We present a method to fit an
Integrate-and-Fire-type model of neuronal activity, namely a
modified version of the Spike Response Model, to a detailed
Hodgkin-Huxley-type neuron model driven by stochastic spike
arrival. In the Hogkin-Huxley model, spike arrival at the synapse
is modeled by a change of synaptic conductance. For such
conductance spike input, more than $70\%$ of the postsynaptic
action potentials can be predicted with the correct timing by the
Integrate-and-Fire-type model. The modified Spike Response Model
is based upon a linearized theory of conductance-driven
Integrate-and-Fire neuron.\bigskip\

\hspace{-\parindent}Keywords: conductance injection -
Integrate-and-Fire model - stochastic input - mapping techniques
- predictive power.\bigskip\

\hspace{-\parindent}PACS: 87.10.+e - 87.19.La - 87.17.Nn -
87.17.Aa.
\end{abstract}
\section{Introduction}
The seminal work by Hodgkin and Huxley \cite{Hodgkin52}, on
mathematical description of action potential generation, has led
to a whole series of models that describe in detail the dynamics
of various ionic currents and the effect of the dendritic
architecture on signal integration, see e.g.
\cite{Bower95,Destexhe99,Mainen96,Traub92}. However, the precise
description of neuronal activity involves a large number of
variables, which often prevents a clear understanding of the
underlying dynamics. Hence, a simplified description is desirable
and has been subject of numerous studies (for a review, see
\cite{Gerstner02a,Maass98}). The most popular simplified models
include the Integrate-and-Fire model \cite{Lapicque07}, the
FitzHugh-Nagumo model \cite{Fitzhugh61,Nagumo62} and the
Morris-Lecar model \cite{Morris81}. However, it is not clear if
such simplified models are sufficient to capture the essence of
neuronal dynamics. Indeed, reduced models of neuronal activity,
as opposed to detailed models of the Hodgkin-Huxley-type, are
commonly thought to be too simple to account for the rich firing
behavior of real neurons.

Nevertheless, these highly simplified models have been shown to
yield good predictions when compared to biological data
\cite{Keat01,Rauch03}. In particular, several parameter
estimation and/or optimization techniques have been proposed to
map reduced models to real neurons
\cite{Izhikevich03,Jolivet04a,Keat01,Paninski04,Rauch03}. These
techniques allow to map electrophysiological data from biological
neurons (intracellular or extracellular recordings) onto
simplified models of the Integrate-and-Fire-type. Moreover, such
a mapping could be the starting point of a systematic
classification of cortical neurons in terms of simplified
dynamics. Finally, reduced models have been used extensively and
successfully to model and study analytically the behaviour of
cortical-like networks \cite{Brunel00,Gerstner00}. In other
words, analytical and computational tools are available to go
from a simple description of single cells dynamics to network
dynamics.

In this paper, we review and extend a numerical technique that
allows a systematic mapping of a class of Integrate-and-Fire
neuron models, namely the Spike Response Model
\cite{Gerstner02a,kistler97}, to intracellular recordings of
neuronal activity \cite{Jolivet04a}. While this technique ensures
reliable predictions and good generalization when the target
neuron is driven by a randomly fluctuating {\em current}, it was
shown in a previous study \cite{Jolivet04a} that the
generalization power is poor when the target neuron is driven by
randomly fluctuating synaptic {\em conductances}. Hence, the
purpose of this paper is to propose extensions of our previous
work which addresses this latter problem. Using recent
theoretical results \cite{Richardson04,Rudolph03}, we propose a
generalization of the Spike Response Model so as to make model
parameters input-dependent. This improved version of the SRM is
shown to be very efficient and robust at predicting the spike
train of a detailed Hodgkin-Huxley type neuron model.
\section{Materials and Methods}
\subsection{The Spike Response Model}
We consider a neuron stimulated by stochastic presynaptic spike
arrival. The state of the neuron is characterized by a single
variable $u$, the membrane voltage of the cell at the soma. Let
us suppose that the neuron has fired its last spike at time
$\hat{t}$. At time $t>\hat{t}$, the membrane potential of the cell
is described as:
\begin{equation}
u(t)=\eta(t-\hat{t})+\sum_{i\in
E}\sum_{f}\epsilon^{+}(t-t_i^{f})+\sum_{j\in
I}\sum_{f}\epsilon^{-}(t-t_j^{f}). \label{equ-SRM}
\end{equation}
The last two terms account for the drive by presynaptic neurons
$i$ (respectively $j$) of the excitatory $E$ (respectively
inhibitory $I$) population. $t_i^{f}$ and $t_j^{f}$ denote the
firing time of presynaptic neurons. The $\epsilon$-functions
describe the excitatory ($+$) and inhibitory ($-$) postsynaptic
potentials (EPSPs and IPSPs). $\eta(t-\hat{t})$ describes the
form of the spike itself as well as the after-hyper-polarization
potential, if present. A spike is elicited if the following
threshold condition is satisfied:
\begin{equation}
    \textrm{if }u(t)\ge\vartheta(t)\textrm{ and }\frac{du}{dt}>0\textrm{ then, }\hat{t}=t.
    \label{equ-threshold-cond}
\end{equation}
Note that spiking occurs only if the membrane voltage crosses the
threshold $\vartheta$ from below. The threshold itself can be
taken either as a constant or as time-dependent. In this paper,
we use a dynamic threshold with the form:
\begin{equation}
    \vartheta(t-\hat{t})=\left\{   \begin{array}{ll}
                                        +\infty & \textrm{if }t-\hat{t}\le\gamma_{ref}\\
                                        \vartheta_0+\vartheta_1\exp(-(t-\hat{t})/\tau_{\vartheta}) & \textrm{else}
                                    \end{array}\right.,
    \label{equ-threshold-def}
\end{equation}
where $\gamma_{ref}$ is a fixed absolute refractory period so as
to exclude continuous firing. $\vartheta_0$, $\vartheta_1$ and
$\tau_{\vartheta}$ are parameters that will be chosen to yield
the best fit to a target spike train (see section
\ref{sec-mapping} below). This version of the Spike Response
Model (the one that we use in this paper) is a simplified version
of the full Spike Response Model and has been termed SRM$_0$ (see
\cite{Gerstner02a} for further details). We will use this acronym
to refer to this model.

Eq. (\ref{equ-SRM}) can be restated in the form:
\begin{equation}
u(t)=\eta(t-\hat{t})+\int_0^{+\infty}\epsilon^{+}(s)\,Q^{+}(t-s)\dd
s+\int_0^{+\infty}\epsilon^{-}(s)\,Q^{-}(t-s)\dd s
\label{equ-SRM-conv}
\end{equation}
with:
\begin{equation}
Q^{+}(t)=\sum_{i\in E}\sum_{f}\delta(t-t_i^f);\quad
Q^{-}(t)=\sum_{j\in I}\sum_{f}\delta(t-t_j^f). \label{equ-def-Q}
\end{equation}
For numerical implementation, we will use a discrete version of
$Q$'s:
\begin{equation}
Q^{+,-}_t=\int_t^{t+\Delta t}Q^{+,-}(s)\dd s
\label{equ-def-Q-disc}
\end{equation}
i.e. $Q^+_t$ is the spike count in a time bin of duration $\Delta
t$ in the excitatory presynaptic population and analogously for
$Q_t^-$. The activity (population rate) $A^+$ of the excitatory
presynaptic population is defined as follows:
\begin{equation}
A^{+}(t)=\frac{1}{\Delta t}\,\frac{\int_t^{t+\Delta t}Q^{+}(s)\dd
s}{N^{+}} \label{equ-def-A}
\end{equation}
where $N^{+}$ is the size of the presynaptic population (with a
corresponding definition for the inhibitory population).
\subsection{Mapping the SRM to voltage traces}
\label{sec-mapping}
The mapping technique that we propose has been discussed in
detail elsewhere \cite{Jolivet04a} and we refer interested
readers to this reference. Here, we describe the essentials of
the technique without going into details.

To realize the mapping of the SRM$_0$ to the target neuron, we
proceed in two steps. First, we extract the functions
characterizing the model (EPSP $\epsilon^{+}$, IPSP
$\epsilon^{-}$ and spike shape $\eta$) and second, we choose the
parameters of the dynamic threshold ($\vartheta$) and optimize
them in terms of quality of predictions. To do so, we assume that
we have at our disposal voltage traces of the target neuron as
well as firing times of presynaptic neurons. We also assume that
the input characteristics are kept constant during the recording
of the dataset used for the mapping procedure. We start by
extraction of the spike shape $\eta$. The shape of spikes is
usually highly stereotyped and presents only little variability.
We therefore select one spike train from the dataset and align
all spikes relatively to some arbitrarily chosen initiation
point. The mean trajectory of the spikes yields $\eta$. Detection
and alignment of spikes is realized using a threshold condition
on the first derivative of the membrane voltage. Once we are done
with $\eta$, we extract the shape $\epsilon^{+}$ of an EPSP and
the shape $\epsilon^{-}$ of an IPSP. If we limit ourselves to the
interval between two consecutive spikes of the same spike train
$\hat{t}_k$ and $\hat{t}_{k+1}$, we can rewrite Eq.
(\ref{equ-SRM}) with the notation introduced in Eqs.
(\ref{equ-SRM-conv}) and (\ref{equ-def-Q-disc}) as follows (for
$\hat{t}_k<t<\hat{t}_{k+1}$):
\begin{equation}
u(t)-\eta(t-\hat{t}_k)=\int_0^{+\infty}\epsilon^{+}(s)Q^+(t-s)\dd
s +\int_0^{+\infty}\epsilon^{-}(s)Q^-(t-s)\dd s.
\label{equ-extract}
\end{equation}
It is then possible to find the optimal $\epsilon$-functions
using the {\em Wiener-Hopf optimal filtering technique}
\cite{Lee65,Wiener58}. We fit the resulting EPSP $\epsilon^{+}$
with a suitable function, typically a difference of exponentials:
\begin{equation}
\epsilon^{+}(s)=K_{\epsilon}^{+}\,\left(\exp(-s/\tau_r^{+})-\exp(-s/\tau_d^{+})\right),
\label{equ-epsilon-kernels}
\end{equation}
with a corresponding definition for $\epsilon^{-}$.
$K_{\epsilon}$, $\tau_r$ and $\tau_d$ are free parameters. The
final step is to choose and optimize the threshold. The absolute
refractory period $\gamma_{ref}$ is set to $2\,$ms. The other
parameters of Eq. (\ref{equ-threshold-def}), i.e. $\vartheta_0$,
$\vartheta_1$ and $\tau_{\vartheta}$, are fitted in order to
optimize the coincidence factor $\Gamma$ (see below) on a given
spike train. In order to ensure a good generalization, we
optimize the threshold with a large dataset generated with
different input characteristics. The numerical optimization
algorithm that we use is the {\em downhill simplex method}
\cite{Nelder65}. Obviously, the SRM$_0$ can only predict neuronal
activity of the specific neuron it has been mapped to.
\subsection{Target neuron model}
Instead of real data from experiments, we use as a reference or
``target'' a Hodgkin-Huxley-type model of a fast-spiking
interneuron \cite{Erisir99}. It contains standard Na$^+$ and
K$^+$ spike-generating channels and one extra Kv$1.3$
K$^+$-channel derived from ``n''-type currents measured in human
T-lymphocytes. This latter current produces a subtle form of
adaptation. See \cite{Erisir99} for further details.
\subsection{Input scenario}
The Hodgkin-Huxley-type model of a fast-spiking interneuron is
driven by random synaptic conductances generated by massive
stochastic presynaptic spike arrival. The total driving current
is given by \cite{Robinson93}:
\begin{equation}
\label{equ-syncurrent} I_{\rm
syn}(t)=g^{+}(t)\,(u(t)-E_+)+g^{-}(t)\,(u(t)-E_-),
\end{equation}
where $g_{+,-}$ are the excitatory ($+$) and inhibitory ($-$)
conductances and $E_{+,-}$ are the corresponding reversal
potential. Synaptic conductances are generated by slightly
correlated homogeneous Poisson spike trains. The excitatory and
inhibitory populations contain respectively $N^+=8000$ and
$N^-=2000$ neurons of which only a fraction $\bar{N}^+/N^+$
(respectively $\bar{N}^-/N^-$) are independent. $\bar{N}^{+,-}$
depend on the presynaptic discharge frequency, on the correlation
coefficient and on the size of the presynaptic population. In the
following, we assume that the presynaptic neurons discharge at
frequencies $\nu^{+}$ and $\nu^{-}$ and the correlation
coefficient is held at a constant value $c=0.002$ (see Appendix
\ref{app-presyn-spikes} for further details about how the
presynaptic spike trains are generated and, in particular, see
Eqs. (\ref{equ-theory-corr-5}) and (\ref{equ-theory-corr-6}) for
the relation between $\bar{N}^{+,-}$ and the correlation
coefficient $c$).

The dynamics of each excitatory synapse $i$ is described by a
conductance variable $P_i^+$ with:
\begin{equation}
\tau_{\rm
syn}^+\dot{P}^+_i=-P^+_i+D^+\sum_f\delta(t-t^f_i)=\frac{D^+}{\tau_{\rm
syn}^+}\,\int_0^{+\infty}e^{-s/\tau_{\rm syn}^+}\,Q^+(t-s)\dd s.
\end{equation}
The value of $P_i^+$ is increased by an amount $D^+$ for each
presynaptic spike activating the synapse at time $t^f_i$. It then
decays back to zero with a time constant $\tau_{\rm syn}^+$. The
total excitatory conductance is the sum of conductance variables
$P_i^+$ over all excitatory synapses. Thus, the total excitatory
conductance is:
\begin{equation}
g^{+}_{\rm syn}(t)=\sum_{i\in E}P^{+}_i(t),
\end{equation}
with a corresponding definition for the total inhibitory
conductance $g^{-}_{\rm syn}(t)=\sum_{j\in I}P^{+}_j(t)$.
Numerical values used in this paper are summarized in Table
\ref{table-syn}.
\begin{table}
    \begin{center}
        \begin{tabular}{cccc}
            \hline\hline
            synapse & $E$ (mV) & $\tau_{\rm syn}$ (ms) & $D$ (mS/cm$^2$)\\
            \hline
            exc. ($+$) & 0 & $2.45$ & $0.073$\\
            inh. ($-$) & -80 & $6.11$ & $0.04$\\
            \hline
        \end{tabular}
    \end{center}
    \caption{Parameters of excitatory and inhibitory synapses (adapted from Destexhe and Par\'e \cite{Destexhe99}). Parameter $D$
    has been adjusted so as to yield an amplitude of postsynaptic potential of the order of $1\,$mV.}
    \label{table-syn}
\end{table}
\subsection{Assessing the quality of predictions of the reduced model}
In order to evaluate quantitatively the predictions of our
reduced model, we use the coincidence factor $\Gamma$ between two
spike trains \cite{kistler97} defined as:
\begin{equation}
\label{equ-gamma} \Gamma=\frac{N_{\rm coinc}-\langle N_{\rm
coinc}\rangle}{\frac{1}{2}(N_{\rm target}+N_{\rm SRM
})}\,\frac{1}{\NN},
\end{equation}
where $N_{\rm target}$ is the number of spikes in the target spike
train $S_{\rm target}$, $N_{\rm SRM}$ is the number of spikes in
the spike train $S_{\rm SRM}$ that is predicted by our reduced
model, $N_{\rm coinc}$ is the number of coincidences with
precision $\Delta$ between the two spike trains, and $\langle
N_{{\rm coinc}}\rangle=2\nu\Delta N_{\rm target}$ is the expected
number of coincidences generated by a homogeneous Poisson process
with the same rate $\nu$ as the spike train $S_{\rm SRM}$. The
factor $\NN=1-2\nu\Delta$ normalizes $\Gamma$ to a maximum value
of one which is reached if and only if the spike train of the
reduced model reproduces exactly that of the target neuron. A
homogeneous Poisson process with the same frequency as the reduced
model model would yield, on average, $\Gamma=0$. We compute the
coincidence factor $\Gamma$ by comparing the two complete spike
trains, i.e., the spike train $S_{\rm target}$ generated by the
target neuron and the train $S_{\rm SRM}$ predicted by the SRM.
This is different to the approach of Kistler and colleagues
\cite{kistler97} where $\Gamma$ was used to predict the {\em
next} spike in a spike train, under the assumption that past
action potentials were correctly reconstructed.
\subsection{Linearized theory of a conductance-driven Integrate-and-Fire model}
The results of section $3.1$ show that a SRM$_0$ with fixed time
course of EPSPs $\epsilon^{+}$ and IPSPs $\epsilon^{-}$ has a
rather limited range of validity. The naive solution would
therefore be to use a different set of EPSPs and IPSPs for each
set of discharge frequencies $\left\{\nu^+;\nu^-\right\}$.
However, this is not a very practical solution if we have to
compute PSPs $\epsilon^{+}$ and $\epsilon^{-}$ for each possible
combination of input frequencies with the method indicated above.
Instead, we aim at finding a simple parameterization of the EPSPs
and IPSPs that would allow to interpolate between and generalize
beyond the specific inputs used for the numerical derivation of
$\epsilon^+$ and $\epsilon^-$. To do so, we use a linearized
theory of conductance-driven Integrate-and-Fire models that
allows to write down an analytical expression for the EPSPs
$\epsilon^{+}$ and IPSPs $\epsilon^{-}$ in function of the
discharge frequencies $\nu^+$ and $ \nu^-$. Thus, the extended
Spike Response Model SRM$_c$ that we derive in this paper is
directly related to the subthreshold dynamics of a
conductance-driven Integrate-and-Fire neuron model (CIF)
\cite{Burkitt03,Richardson04,Rudolph03}.

The subthreshold membrane voltage $u$ of a CIF neuron model is
given by the following differential equation:
\begin{equation}
C\,\frac{d}{dt}u=-g_L\,(u-E_L)-g^+(t)\,(u-E^+)-g^-(t)\,(u-E^-),
\label{equ-CIF-def}
\end{equation}
with $C$ the membrane capacitance, $g_L$ the leak conductance
(with a reversal potential $E_{L}$) and $g^{+}$ (respectively
$g^{-}$) the instantaneous excitatory (respectively inhibitory)
conductance. $E^{+}$ and $E^{-}$ are the reversal potentials of
the excitatory and inhibitory synapses. It is straightforward to
show that the average membrane voltage predicted by this model (in
absence of spiking mechanism) is given by:
\begin{equation}
\mu_{\rm CIF}=\frac{g_L\,E_L+\bar{g}_{\rm
syn}^+\,E_++\bar{g}_{\rm syn}^-\,E_-}{g_L+\bar{g}_{\rm
syn}^++\bar{g}_{\rm syn}^-}.\label{equ-muCIF}
\end{equation}
The bars denote time averaging in this case. The EPSPs
$\epsilon^{+}$ and IPSPs $\epsilon^{-}$ of the SRM$_0$ that would
correspond to the CIF model can then be calculated as the linear
response around the average membrane voltage $\mu_{\rm CIF}$. We
find:
\begin{equation}
\epsilon^{+}(s)=\frac{D^+\,\tau_{\rm eff}\,\tau_{\rm
syn}^+\,(\mu_{\rm CIF}-E^+)}{\tau_{\rm eff}-\tau_{\rm
syn}^+}\,\left(e^{-s/\tau_{\rm syn}^{+}}-e^{-s/\tau_{\rm
eff}}\right). \label{equ-PSPs-CIF}
\end{equation}
with a corresponding definition for $\epsilon^{-}(s)$. We compare
the results of Eqs. (\ref{equ-muCIF}) and (\ref{equ-PSPs-CIF})
with results extracted from simulations of the target
Hodgkin-Huxley-type neuron model. We find that, outside the regime
of EPSP-amplification, both the average membrane voltage (data
not shown) and the shape of PSPs $\epsilon^{+}$ and
$\epsilon^{-}$ are well predicted by this simple linearized
theory (see Figure \ref{fig-kernels}A and B in \textit{Results}
section). To illustrate this, we fit the PSPs plotted in Figure
\ref{fig-limitations}A with Eq. (\ref{equ-PSPs-CIF}) with
$\tau_{\rm eff}$ as a free parameter. We then compare this
parameter with $\tau_{\rm eff}$ as predicted by the Eq.
(\ref{equ-taueff}). Figure \ref{fig-taueff} shows that the
linearized theory is in very good agreement with the results of
simulations of the target Hodgkin-Huxley-type neuron model except
in the area where EPSP-amplification takes place.

Equation (\ref{equ-PSPs-CIF}) gives an analytical expression for
the PSPs $\epsilon^+$ and $\epsilon^-$. The PSPs allow us to
reproduce the fluctuations of the membrane voltage. However, we
also need to account for the constant voltage bias which appears
when discharge frequencies are elevated. Therefore, we redefine
the first term of the SRM$_0$, namely the function $\eta$ as:
\begin{equation}
\eta(t-\hat{t})\to\eta(t-\hat{t})+(\mu_{\rm
CIF}-E_L)\label{equ-eta-CIF}.
\end{equation}
This simple procedure ensures that our model produces the correct
average membrane voltage. The model could be further improved by
using a time-dependent leak conductance $g_L(t-\hat{t})$
\cite{Jolivet04a}, but we will not do so.

In order to relate the linearized theory of conductance-driven
Integrate-and-Fire model of section $3.2$ to the numerical PSPs
$\epsilon^+$ and $\epsilon^-$ extracted in subsection $3.1$, we
need to estimate the parameters appearing in Eq.
(\ref{equ-PSPs-CIF}) from the data. We will assume that the size
of the presynaptic populations ($N^+$ and $N^-$) as well as the
average discharge frequencies in these populations ($\nu^+$ and
$\nu^-$) are known. We will also assume ``standard'' reversal
potentials at the synapses, i.e. $E^+=0\,$mV and $E^-=-80\,$mV.
$C$ is taken to be constant at a value of $1\,\mu$F/cm$^2$. The
reversal potential of the leak current can be crudely
approximated by the resting potential of the neuron $E_L\approx
u_{\rm rest}$.

Thus, the parameters we need to estimate are the synaptic time
constants $\tau_{\rm syn}^{+}$ and $\tau_{\rm syn}^{-}$, the
effective membrane time constant $\tau_{\rm eff}$, as well as the
mean conductances $g_L$, $\bar{g}^+$ and $\bar{g}^-$. The time
constants $\tau_{\rm syn}^{+}$ (respectively $\tau_{\rm
syn}^{-}$) can be estimated by fitting the EPSP $\epsilon^{+}$
(respectively the IPSP $\epsilon^{-}$) extracted from a dataset
where EPSP-amplification doesn't take place, i.e. a dataset with
strong inhibition and weak excitation. Once $\tau_{\rm syn}^{+}$
and $\tau_{\rm syn}^{-}$ are known, one can find the other
parameters by comparing EPSPs $\epsilon^{+}$ extracted for
several different sets of input characteristics. Fitting the
EPSPs $\epsilon^{+}$ with formula (\ref{equ-PSPs-CIF}) yields an
estimate of $\tau_{\rm eff}$. If we collect three different
values of $\tau_{\rm eff}$ for three different sets of input
characteristics ($\nu^{+}$ and $\nu^{-}$) and using the
definition of $\tau_{\rm eff}$ (see Eq. (\ref{equ-taueff}) in
\textit{Results} section), we find a set of three equations with
three unknowns, namely $g_L$, $\bar{g}_{\rm syn}^+$ and
$\bar{g}_{\rm syn}^-$. $\bar{g}_{\rm syn}^+$ is given by:
\begin{equation}
\bar{g}_{\rm syn}^+=N^+\,\nu^{+}\,\tau_{\rm syn}^+\,D^+
\end{equation}
with a corresponding definition for $\bar{g}_{\rm syn}^-$.
Solving this set of equations thus yields an estimate of $g_L$,
$D^+$ and $D^-$. Therefore, we now have a simple model that
allows to compute PSPs $\epsilon^{+}(s)$ and $\epsilon^{-}(s)$ in
a straightforward fashion for any given set of input discharge
frequencies $\nu^{+}$ and $\nu^{-}$. We will refer to this model
as SRM$_c$ with a lower-case ``c'' for {\em conductance-based}.
\section{Results}
In the first subsection, we discuss the intrinsic limitations of
the simple Spike Response Model (SRM$_0$). In particular, we show
that the shape of EPSPs and IPSPs derived by our method changes
systematically as a function of the input parameters. These
findings are then compared with a linearized theory of
conductance-driven Integrate-and-Fire models (CIF). This
comparison allows us to determine the parameters of a new
conductance-based Spike Response Model (SRM$_c$) which is tested
over a broad range of different inputs in subsection $3.2$.
\subsection{Limitations of a classic SRM$_0$}
We map the target neuron model to the SRM$_0$ using the technique
described in {\em Materials and Methods}. Let us recall that the
SRM$_0$ is characterized mainly by the spike shape $\eta$ as well
as the EPSP $\epsilon^+$ and the IPSP $\epsilon^-$. Fig.
\ref{fig-limitations}A shows the EPSPs and IPSPs $\epsilon^{+,-}$
extracted for different input discharge frequencies $\nu^{+}$ and
$\nu^{-}$ and Fig. \ref{fig-limitations}B shows the spike shape
$\eta$. The shape of the spike does not depend on the
characteristics of the input scenario. One immediately remarks
that both the characteristic time scales and the amplitude of the
EPSP/IPSP do change in function of the input discharge
frequencies. In fact, the numerical technique extracts the best
linear filters so as to reproduce the membrane voltage trace of a
given sample spike train. The corresponding EPSPs and IPSPs are
then optimal for the specific set of input discharge frequencies
used for parameter extraction but there is no reason why they
should be optimal for other sets of inputs with different
characteristics. Indeed, there are reasons why they should be
different depending on input discharge frequencies. Let's quickly
review these reasons \cite{Destexhe99}.

First, two different sets of input discharge frequencies are
likely to produce two different values of average membrane
voltage. In turn, this means increased or decreased average
driving forces for the synapses as the corresponding current
includes a multiplicative term of the form $(u(t)-E^{+,-})$ (see
Eq. (\ref{equ-syncurrent})). This should affect the amplitude of
the EPSPs and IPSPs. Second, two different sets of input discharge
frequencies are likely to produce two different total
conductances. This affects the effective membrane time constant
of the neuron. If we neglect the effect of somatic AP-generating
ion channels, the effective time constant in the subthreshold
regime can be written:
\begin{equation}
\tau_{\rm eff}=\frac{C}{g_L+\bar{g}_{\rm syn}^++\bar{g}_{\rm
syn}^-}, \label{equ-taueff}
\end{equation}
with $C$ the membrane capacitance, $g_L$ the conductance of the
leak current and $\bar{g}_{\rm syn}^+$ and $\bar{g}_{\rm syn}^-$
the total average excitatory and inhibitory synaptic
conductances. Finally, when the excitatory input discharge
frequency is very high, the target neuron runs in a highly
non-linear regime due to activation of Na$^+$-channels. This
effect is known as EPSP-amplification \cite{Fricker00} and
illustrated for our target neuron in Fig. \ref{fig-limitations}C
and D. While EPSP-amplification is usually not observed in
fast-spiking interneurons, it is present in our target model of a
fast-spiking interneuron since Na$^+$-channels are the only
channels opened at depolarized states close to threshold (see
{\em Materials and Methods} section). All these three effects
(average driving force, effective membrane time constant and
EPSP-amplification) combine with each other and lead to the
pattern of EPSPs and IPSPs shown in Fig. \ref{fig-limitations}A
and C. The EPSPs and IPSPs are shorter when total conductance is
increased (shortening of effective membrane time constant) except
when this increase is mainly due to the excitatory conductance
which then leads to the reverse effect because of
EPSP-amplification (see panel C). The amplitude of EPSPs
decreases when excitation is increased and increases when
excitation is decreased (average driving force) with
corresponding effects for IPSPs and modifications of the
inhibition level. Interestingly, one remarks that when inhibition
is very strong together with weak excitation, the extracted EPSPs
$\epsilon^+$ and IPSPs $\epsilon^-$ follow exactly the dynamics
of the corresponding synapses, i.e. an almost instantaneous rise
followed by a decay with the same time constant as the time
constant of the synaptic conductance (see Fig.
\ref{fig-limitations}A at bottom right). The dynamics of EPSPs
and IPSPs can be approximated by:
\begin{figure}[!htp]
    \centering\includegraphics[width=0.8\linewidth]{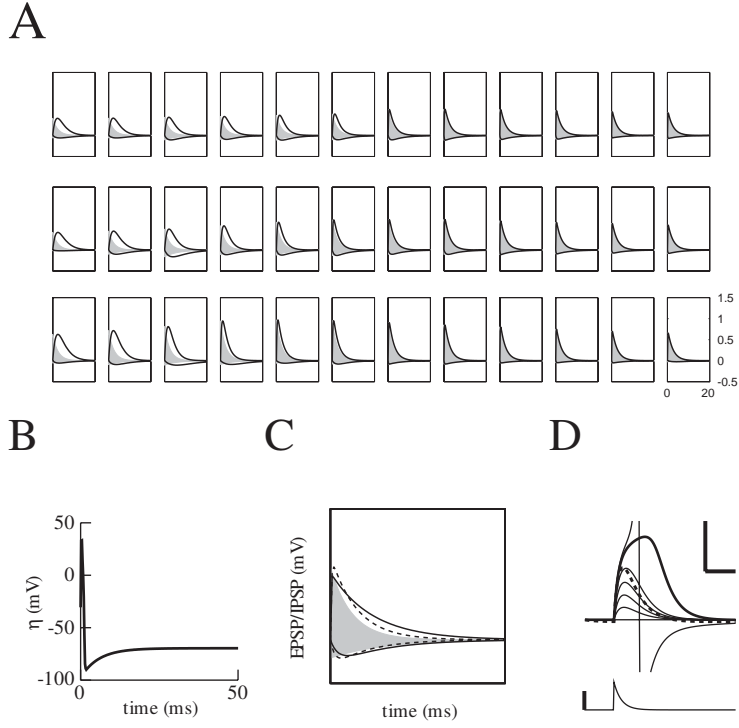}
    \caption[]{EPSPs, IPSPs and spike shape as extracted by our numerical method. {\bf A.} EPSPs and IPSPs (respectively positive and negative solid lines) for
    different presynaptic input discharge frequencies $\nu^{+}=0.9,0.6,0.3\,$Hz (from top to bottom) and $\nu^{-}=1,2,3,\cdots,12\,$Hz (from left to right). Horizontal
    axis is in ms and vertical axis is in mV. The light grey area shows the dynamics of the normalized EPSCs and IPSCs (vertical axis in arbitrary units). {\bf B.} The
    spike shape $\eta$ {\bf C.} Effect of Na$^+$-channels on the shape of kernels. This panel repeats the results of panel A with discharge frequencies $\nu^{-}=5\,$Hz
    and $\nu^{+}=0.9\,$Hz (light grey area and solid lines). The dotted lines correspond to EPSPs and IPSPs extracted while blocking Na$^+$-channels. Resulting PSPs
    are significantly shorter. {\bf D.} EPSP-amplification in the target model. An exponentially decaying excitatory conductance is injected in the target neuron
    (bottom; vertical bar is $0.005\,$mS/cm$^2$). The response of the target neuron (top) when receiving $n$-fold times the excitatory conductance for
    $n=0,4,8,12,16,20$ (thin solid lines). For $n=20$, one observes huge EPSP-amplification in normal conditions (thick solid line) but none when Na$^+$-channels are
    blocked (thick dotted line; note that in this latter case, the resting state is slightly changed) and a spike is finally elicited for $n=21$. Horizontal bar is $10\,$ms
    and vertical bar is $10\,$mV.}
    \label{fig-limitations}
\end{figure}
\begin{equation}
\epsilon^{+,-}(s)\propto \left(e^{-s/\tau_{\rm
syn}^{+,-}}-e^{-s/\tau_{\rm eff}}\right). \label{equ-EPSP-approx}
\end{equation}
Therefore, when drive is very strong, $\tau_{\rm eff}\approx 0$
from Eq. (\ref{equ-taueff}) and $\epsilon^{+,-}$ is reduced to
$e^{-s/\tau_{\rm syn}^{+,-}}$. Note that this effect does appear
only with weak excitatory stimulation. When excitatory drive is
strong, EPSP-amplification compensates the reduction of
$\tau_{\rm eff}$ and Eq. (\ref{equ-taueff}) does not hold anymore.

Given the change of the time course of EPSPs and IPSPs as a
function of the input, it is clear that correct predictions of the
subthreshold fluctuations of the membrane voltage by linear
summation of EPSPs and IPSPs with a fixed time course are only
possible in the input regime for which the reduced model was
constructed (recall Eq. (\ref{equ-SRM})).

In order to make the EPSPs and IPSPs input dependent, we
parameterize $\epsilon^+$ and $\epsilon^-$ using the linearized
theory of a conductance-driven Integrate-and-Fire model
\cite{Burkitt03,Richardson04,Rudolph03}; see \textit{Materials
and Methods} for details.

\begin{figure}[!htp]
    \centering\includegraphics[width=0.6\linewidth]{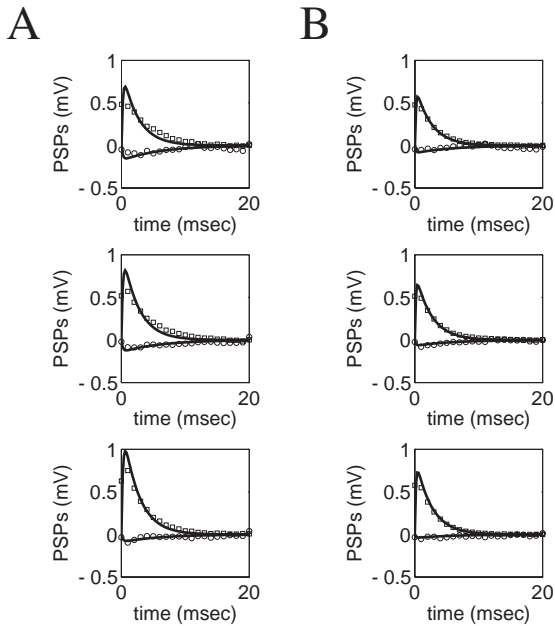}
    \caption[]{The EPSPs and IPSPs as predicted by the linearized theory (solid lines; see Eq. (\ref{equ-PSPs-CIF})) are compared
    to the EPSPs and IPSPs extracted by the method proposed in \textit{Materials and Methods} (symbols). {\bf A. } The inhibitory
    discharge frequency $\nu^-=6\,$Hz and the excitatory discharge frequency $\nu^+=0.9,\,0.6,\,0.3$ (from top to bottom). {\bf B. }
    Same as in A except that $\nu^-=10\,$Hz.}
    \label{fig-kernels}
\end{figure}

\begin{figure}[!htp]
    \centering\psfig{figure=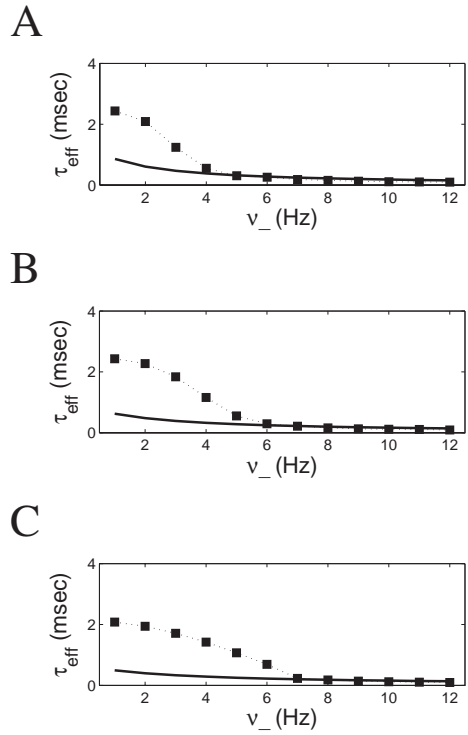,width=0.5\linewidth}
    \caption[]{The effective membrane time constant as predicted by a CIF (solid
    line) is compared to the corresponding parameter extracted from simulations
    of the target Hodgkin-Huxley-type neuron model (dotted line with squares).
    See text for further details. {\bf A.} $\nu^{+}=0.3\,$Hz. {\bf B.}
    $\nu^{+}=0.6\,$Hz. {\bf C.} $\nu^{+}=0.9\,$Hz.}
    \label{fig-taueff}
\end{figure}

We find that outside the regime of EPSP-amplification the
numerically derived postsynaptic potentials $\epsilon^+$ and
$\epsilon^-$ are well fitted by the theory (see Figure
\ref{fig-kernels}). The regime of EPSP-amplification can easily
be identified by comparing the effective membrane time constant
$\tau_{\rm eff}$ predicted by the theory with that derived from
the numerically optimized PSPs $\epsilon^+$ and $\epsilon^-$ (see
Figure \ref{fig-taueff}). All parameters of the linearized
theory, in particular the synaptic time constants $\tau_{\rm
syn}^+$, $\tau_{\rm syn}^-$ and the mean conductances $\bar{g}^+$
and $\bar{g}^-$ can hence be estimated from the data by using
three sets of inputs that do not lead to EPSP-amplification (see
\textit{Materials and Methods}). The resulting model is a
conductance-driven Spike Response Model (SRM$_c$) which we now
test on new set of inputs independent from the one used for
parameter optimization.
\subsection{Predicting spike by spike}
A conductance-driven Spike Response Model (SRM$_c$) has been
estimated from the numerical voltage traces using the procedure
described in the \textit{Materials and Methods} section. We now
test the predictive power of the SRM$_c$. We are interested at
reproducing the exact timing of the spikes of the target neuron.
As we have not dealt yet with the threshold, the first step is to
optimize the three free parameters of the threshold, namely
$\vartheta_0$, $\vartheta_1$ and $\tau_{\vartheta}$ (see Eq.
(\ref{equ-threshold-def}), $\gamma_{ref}$ is set at a constant
value of $2\,$ms). To do so, we use a very long spike train with
input characteristics that include the discharge frequencies
where EPSP-amplification doesn't take place (see Figure
\ref{fig-taueff}). We then proceed as indicated in
\textit{Materials and Methods}. The resulting numerical values
are summarized in Table \ref{table-thresh}.
\begin{table}[!ht]
    \begin{center}
        \begin{tabular}{cccc}
            \hline\hline
            parameter & mean & SD\\
            \hline
            $\vartheta_0$ (mV)      & $-38.437$ & $0.002$\\
            $\vartheta_1$ (mV)      & $564.0$ & $0.7$\\
            $\tau_{\vartheta}$ (ms) & $0.91$ & $0.03$\\
            \hline
        \end{tabular}
    \end{center}
    \caption{Best fit parameters for the threshold (see text for further details).
    Mean and standard deviation (SD) are computed from four optimizations with
    different initial conditions. The small standard deviation shows that the four
    optimization runs all converge to the same minimum.}
    \label{table-thresh}
\end{table}

\begin{figure}[!t]
    \centering\psfig{figure=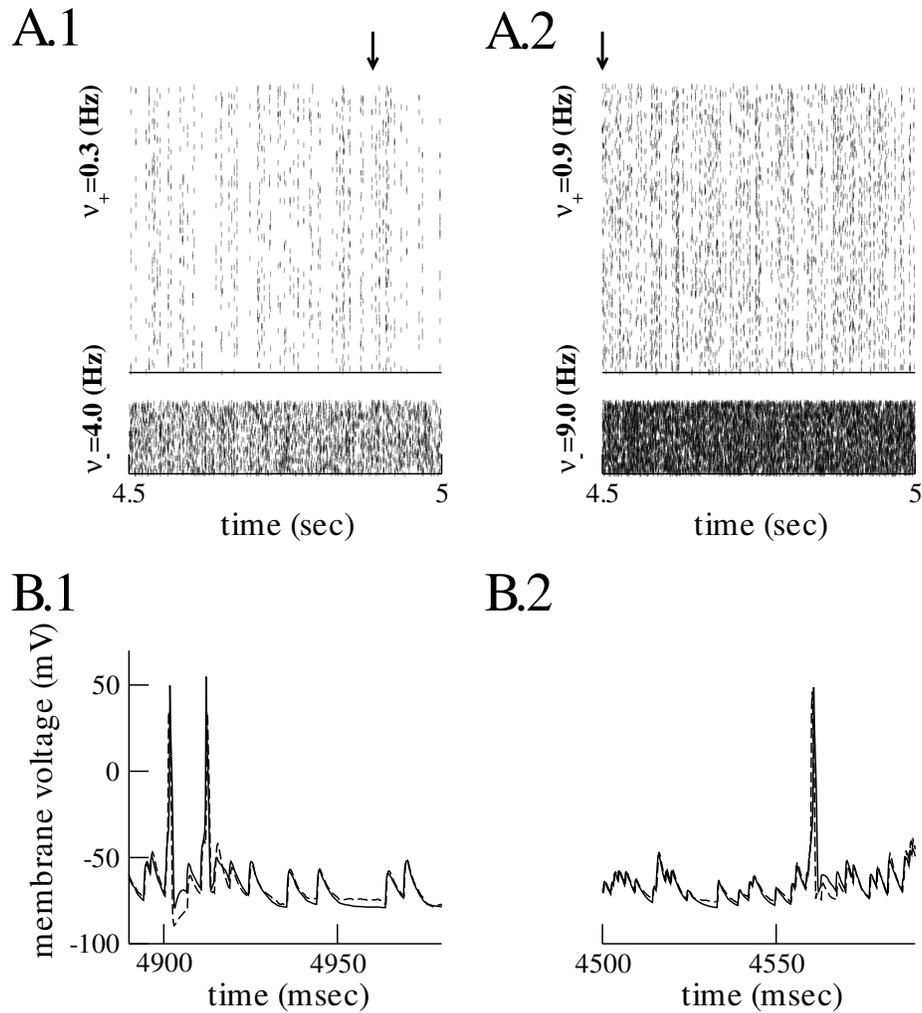,width=1.0\linewidth}
    \caption[]{Comparative results for low (1) and high (2) input discharge frequencies.
    {\bf A.} The activity in the excitatory (top) and inhibitory (bottom) presynaptic
    populations. The arrows indicate the starting point of the segment plotted in B. {\bf B.}
    Corresponding membrane voltage of the target neuron (solid line) is compared to the membrane voltage
    as predicted by the SRM$_c$ (dashed line). In both cases (B.1 and B.2) the membrane voltage as predicted by
    the SRM$_c$ gives a fair approximation of the membrane voltage of the target neuron.}
    \label{fig-comparative}
\end{figure}

Figure \ref{fig-comparative} shows sample results of the
predictive power of the SRM$_c$ in two distinct regimes,
low-drive (low presynaptic discharge frequencies, 1) and
high-drive (high presynaptic discharge frequencies, 2). In both
cases the predicted membrane voltage is reasonably close to the
membrane voltage of the target neuron. Note also that all the
spikes in plotted samples are reproduced with the exact timing
(panel B). For these two cases, we find $\Gamma=0.76$ (low-drive)
and $\Gamma=0.67$ (high-drive). The coincidence factor $\Gamma$
takes a value of $1$ if $100\%$ of spikes coincide and is
normalized to $0$ if coincidences are random (see {\em Materials
and Methods} section).

\begin{figure}[!ht]
    \centering\psfig{figure=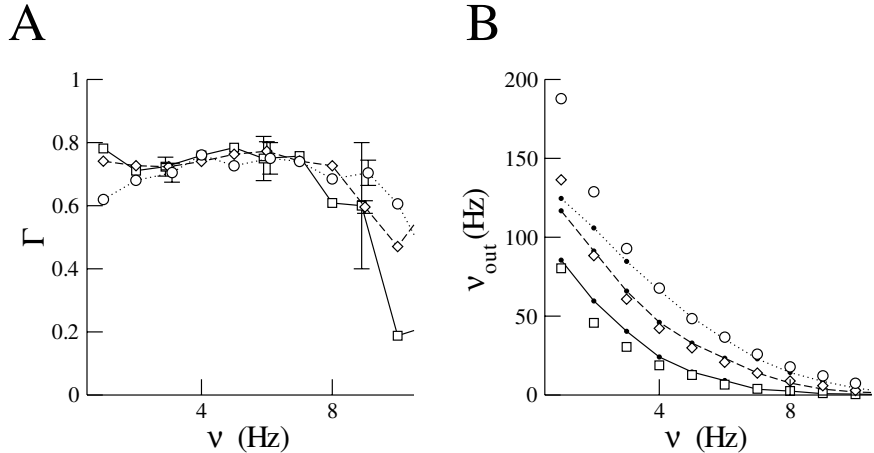,width=1.0\linewidth}
    \caption[]{Coincidence factor $\Gamma$ (A; mean $\pm$ SD) and output frequency $\nu_{\rm out}$ (B) plotted versus
    the inhibitory input discharge frequency $\nu^{-}$ for various values of the excitatory input discharge frequency
    $\nu^{+}$ ($\nu^{+}=0.3\,$Hz solid line and squares; $\nu^{+}=0.6\,$Hz dashed line and diamonds; $\nu^{+}=0.9\,$Hz dotted line and circles).
    In the panel B, the output frequency of the SRM$_c$ (symbols) is compared to the output frequency of the target neuron (dots and line).}
    \label{fig-gamma-f}
\end{figure}

To test the performances in a more systematic way, we quantify the
predictions of the SRM$_c$ in terms of the timing of the spikes
(coincidence factor $\Gamma$) and in terms of output frequencies
of the model ($\nu_{\rm out}$) over a broad range of input
characteristics ($\nu^{+}$ and $\nu^{-}$). Figure
\ref{fig-gamma-f} shows the performances of the SRM$_c$ for such
a systematic procedure. We observe that the SRM$_c$ yields good
performances ($\Gamma\ge 0.7$ and output frequency $\nu_{\rm
out}$ predicted in the correct range) over a broad range of input
discharge frequencies. the only notable exception is when
inhibitory discharge frequency is very high. In this regime, the
output frequency $\nu_{\rm out}$ is close to zero so that the
number of spikes in a spike train is low. Therefore, our
coincidence factor $\Gamma$ is not well suited for this case.
However, we note that even if $\Gamma$ is low, the output
frequency is predicted in the correct range. Furthermore, the
subthreshold fluctuations of the membrane voltage are well
reproduced (not shown).

Interestingly, the SRM$_c$ yields good performances where it is
supposed to do so but also where EPSP-amplification takes place
(see Figure \ref{fig-taueff}). In the case of our target neuron,
EPSP-amplification occurs when both excitatory and inhibitory
populations discharge at rather low frequencies. A fixed
coefficient of correlation $c$ then imposes co-activation of large
subpopulations of synapses in this regime (see Eq.
(\ref{equ-theory-corr-5})). The target neuron therefore spends
most of the time close from the resting potential and
occasionally receives numerous coincident presynaptic spikes that
bring it in the area of EPSP-amplification. Most of these large
excursions then lead to a spike while only a few of them do not.
In this situation, one can easily predict spikes at a correct
timing with a threshold model like the SRM$_c$.

In summary, our modified version of the SRM$_0$, the so-called
SRM$_c$, performs well over a broad range of input characteristics
$\nu^{+}$ and $\nu^{-}$ and predicts not only the output
frequency but also most of the spikes with the correct timing. It
is useful to keep in mind that we used a calculation based on a
conductance-driven Integrate-and-Fire model to evaluate the
subthreshold membrane voltage fluctuations and the average
membrane voltage. This latter results do not include the
potential effect of ionic channels activated in the subthreshold
regime. The SRM$_c$ should therefore fail in regimes where such
ionic channels are activated. However, we found here that even
though the underlying theoretical framework is on the edges of
its validity domain, the SRM$_c$ still performs reasonably well
in the regime where EPSP-amplification takes place and thus as
long as there are large enough fluctuations of the subthreshold
membrane voltage.
\section{Discussion}
Mapping real neurons to simplified neuronal models has benefited
from many theoretical developments in recent years and applied to
both {\em in-vitro} and {\em in-vivo} recordings
\cite{Keat01,Rauch03}. However, most of the techniques have been
developed for a current injection scenario
\cite{Brillinger79,Brillinger88a,Paninski04,Rauch03}. On the
experimental side, conductance injection is increasingly used
instead of current injection and is thought to be closer to {\em
in-vivo} conditions (see \cite{Destexhe03} for a review).

We had previously reported a mapping technique \cite{Jolivet04a}
based on standard signal processing tools which allows a
systematic mapping of a simplified neuron model, the Spike
Response Model \cite{Gerstner02a,kistler97}, to intracellular
recordings. It has been shown to yield very good results in the
case of current injection for model neurons \cite{Jolivet04a} and
with real data (unpublished observations). However, while the
reduced model built in this way generalizes its predictions over
a broad range of different input characteristics for the current
injection case, it performs very poorly in the conductance
injection case for reasons reviewed above.

In this paper, we have shown that a simple modification of the
classic SRM$_0$ solves this problem. In its new formulation, the
model is able to predict very reliably many aspects of neuronal
activity, such as timing of the spikes, membrane voltage and mean
output rate. The global performances are improved and moreover,
the model can generalize predictions extremely well. The new
model SRM$_c$ is directly related to conductance-driven
Integrate-and-Fire neurons
\cite{Burkitt03,Richardson04,Rudolph03}. Our technique can be
applied to extract simple neuron models from experimental
intracellular recordings under conductance injection.
\begin{appendix}
\newcommand{\nn}{{\mathcal{N}}}
\section{Generation of presynaptic spike trains}
\label{app-presyn-spikes}
In this appendix, we detail the method used in simulations to
generate slightly correlated spike trains and we also derive some
useful analytical results. The method follows \cite{Destexhe99}
but there exists other ways to generate correlated spike trains
(see \cite{Kuhn03} for instance).

Presynaptic spike trains are described by random homogeneous
Poisson processes. At each time step, $\bar{N}$ independent
random variables are generated and distributed among the
$N\ge\bar{N}$ presynaptic neurons to generate slightly correlated
spike trains \cite{Destexhe99}. In this specific case, we can
derive the probability distribution function (PDF) of the
variable $Q$ (see Eq. (\ref{equ-def-Q-disc}) in \textit{Materials
and Methods} section for a definition of variable $Q$). Here, all
the calculations rely on a discrete time scale with bins of width
$\Delta t$. Let us consider that elements of a vector $\bar{V}$
of length $\bar{N}$ are distributed at random and receive either
a value of $1$ with a constant probability $p$ or $0$ with
probability $(1-p)$. In this specific case, $p=\nu\,\Delta t$
with $\nu$ the average discharge frequency in the presynaptic
population and $\Delta t$ the size of the time steps used in the
simulation. The total number $K$ of elements receiving a value of
$1$ in $\bar{V}$ is therefore distributed according to a binomial
distribution $P(K=k)=B(k;\bar{N},p)$. In a second vector $V$ of
length $N$, elements receive $0$ or $1$ according to a parent
element chosen at random in vector $\bar{V}$. The probability of
receiving a value of $1$ is then $\tilde{p}=K/\bar{N}$. The total
number of elements $Q$ of $V$ receiving a value of $1$ is then
given by $P(Q=q|K)=B(q;N,\tilde{p})$. The average distribution of
variable $Q$ is then:
\begin{equation}
\label{equ-theory-PDF-0}
P(Q=q)=\sum_{k=0}^{\bar{N}}P(Q=q|K)P(K=k).
\end{equation}
Some algebra yields:
\begin{equation}
\label{equ-theory-PDF}
P(Q=q)=C_q^N\,\bar{N}^{-N}\,\sum_{k=0}^{\bar{N}}C_k^{\bar{N}}\,p^k\,(1-p)^{\bar{N}-k}\,k^q\,(\bar{N}-k)^{N-q}.
\end{equation}
In the following, we will need to know the first two moments
$E[Q]$ and $Var[Q]$ of this distribution. Using the definition of
$\tilde{p}$ (see above), we find that $E[\tilde{p}]=p$ and thus:
\begin{equation}
\label{equ-theory-mean} E[Q]=N\,p.
\end{equation}
To calculate the variance of $Q$, we use the fact that:
\begin{eqnarray}
E[Q^2]  &=& Var[Q]+E[Q]^2\nonumber\\
        &=& N\tilde{p}\,(1-\tilde{p})+N^2\tilde{p}^2
\end{eqnarray}
Using the fact that $E[\tilde{p}]=p$ and that
$E[\tilde{p}^2]=\bar{N}^{-2}\,Var[K]+p^2$, we find that:
\begin{equation}\label{equ-theory-var}
Var[Q]=Np\,(1-p)\,\left(1+N/\bar{N}\right)
\end{equation}

While variables $N$ and $p$ have direct biological
interpretations ($N$ is the size of the presynaptic population and
$p$ is related to the discharge frequency in that population),
$\bar{N}$ is a rather abstract quantity which is linked to the
correlations in the activity of the presynaptic population. For
practical use, it would be useful to compute the correlation
coefficient between two spike trains in terms of these variables.
The correlation coefficient between two sequences of numbers $m_i$
and $n_i$ is defined by:
\begin{equation}\label{equ-corr-def}
c=\frac{\sum_im_i\,n_i}{\sqrt{\sum_im_i^2\,\sum_jn_j^2}}
\end{equation}
If we choose at random two elements from the vector $V$, the
probability that both fire together is given by the
hypergeometric distribution $H(N,q,2)$ so that:
\begin{equation}\label{equ-theory-corr-1}
P(Y=2|Q)=\frac{C_2^q\,C_{0}^{N-q}}{C_2^{N}}=\frac{q\,(q-1)}{N\,(N-1)}.
\end{equation}
On the other hand, if we choose only one element from the vector
$V$, the probability that it fires is given by $H(N,q,1)$:
\begin{equation}\label{equ-theory-corr-2}
P(Z=1|Q)=q/N
\end{equation}
Finally, the average correlation coefficient between two randomly
chosen spike trains of the presynaptic population is given by:
\begin{equation}\label{equ-theory-corr-3}
c=\frac{\sum_{q=0}^NP(Y=2|Q)\,P(Q=q)}{\sum_{q'=0}^NP(Z=1|Q)\,P(Q=q')}.
\end{equation}
Using Eqs. (\ref{equ-theory-corr-1}) and
(\ref{equ-theory-corr-2}) in Eq. (\ref{equ-theory-corr-3}), we
find:
\begin{equation}\label{equ-theory-corr-4}
c=\frac{1}{N-1}\,\frac{E[Q^2]-E[Q]}{E[Q]}.
\end{equation}
Note that this latter result is general when considering
homogeneous Poisson spike trains with the same rate $p$ and
doesn't depend on the specific way spike trains are generated. In
our case, some algebra yields:
\begin{equation}\label{equ-theory-corr-5}
c=p+\frac{N\,(1-p)}{\bar{N}\,(N-1)}
\end{equation}
and thus:
\begin{equation}\label{equ-theory-corr-6}
\bar{N}=\frac{N\,(1-p)}{(N-1)\,(c-p)}.
\end{equation}
\end{appendix}
\section*{Acknowledgments}
The authors would like to thank Magnus Richardson for fruitful
discussions.
\section*{Grants}
RJ is supported by the {\em Swiss National Science Foundation}
under grant number N FN-2100-065268.
\end{document}